 \documentclass[nohyper,12pt,letterpaper]{JHEP}
 \usepackage{graphicx}
 \usepackage{axodraw}
 \newfont{\frak}{eufm10 scaled 1200}
 
 \newfont{\Bbb}{msbm10 scaled 1200} 
 \newcommand{\mathbb}[1]{\mbox{\Bbb #1}}
 \DeclareSymbolFont{AMSa}{U}{msa}{m}{n}
 \DeclareSymbolFont{AMSb}{U}{msb}{m}{n}
 \let\Box\relax
 \DeclareMathSymbol{\Box}{\mathord}{AMSa}{"03}

 \def \eqn#1#2{\begin{equation}#2\label{#1}\end{equation}}

 \title{Remodeling the Pentagon After the Events of\\ \centerline{{ $\bf 2/23/06$}}}
 \author{T.\,Banks\\
 Department of Physics \\
 University of California, Santa Cruz, CA 95064\\
 E-mail: \email{banks@scipp.ucsc.edu}\\
 {\it and}\\
 NHETC, Rutgers University\\
 Piscataway, NJ 08540}

 \abstract{The meta-stable SUSY breaking mechanism of Intriligator Seiberg and Shih can be used to
 simplify the Pentagon model of TeV scale physics.   The simplified model has only a single scalar field
 and no troublesome low energy axion.   One significant signature is  $l^+ l^-  X$ plus missing energy, where
 $X$ might be the two photons of gauge mediated models, but is likely to be different. There is
 a new strongly interacting sector with a scale around $1.5$ TeV.  The penta-hadrons
 of this sector have masses of order $6$ TeV or more.  Dark matter is probably
 the pseudo-goldstone boson of spontaneously broken penta-baryon number.  This can be a viable dark matter candidate if
 an appropriate asymmetry in penta-baryon number is generated in the early universe.  The pseudo-Goldstone
 particle has a mass of $\sim 1$ eV and is produced predominantly in flavor changing charged current decays of ordinary particles.
 The model solves the flavor problems of SUSY, but has two low energy CP violating phases, whose
value is strongly constrained by experiment.}
 \received{} \accepted{}
 \preprint{\hepph{}\\\\RU-06-07/SCIPP-06-06 \\}
 
\begin{document}
 \section{\bf Introduction}

In a recent paper\cite{pentagon}, I proposed an explicit model,
which implemented the idea of Cosmological SUSY Breaking (CSB).
Apart from the fields of the supersymmetric standard model (SSM),
the model consisted of the Pentagon - a new strongly interacting
$SU(5)$ super-QCD with $5$ flavors of penta-quark, and three singlet
fields, $S,G,T$ with a variety of Yukawa couplings to the Pentagon
and the SSM.   The intricate pattern of singlets was required to
motivate the possibility of a meta-stable SUSY breaking vacuum state
of the flat space quantum field theory.   It also caused a potential
phenomenological problem - a low scale QCD axion.   One could invoke
higher dimension operators, which raised the axion mass and made it
barely compatible with observation.   This of course removed the
model's solution of the strong CP problem.   It also introduced a
fine tuning of the electroweak scale, of order $1\%$.   This was
required to raise the axion decay constant above the laboratory
bounds.

The Pentagon Paper was written before the world changed on
$02/23/06$.   The Neo-conservative\footnote{This term is motivated
by the fact, pointed out to the author by N. Seiberg , that
meta-stable SUSY breaking was advocated in unpublished work by M.
Dine, which was done in the run-up to the Affleck Dine Seiberg
\cite{ads} discovery of models of dynamical SUSY breaking.}
revolution in SUSY breaking, heralded by the paper of Intriligator,
Seiberg, and Shih (ISS)\cite{iss}\ now provides us with a plethora
of calculable SUSY breaking models, where SUSY is broken in a
meta-stable state - exactly what is needed to solve the problems at
the Pentagon.

ISS prove that when a mass term $m_{ISS} {\rm tr} P\tilde{P} $ is
added to SUSY QCD with $N_C +1 \leq N_F \leq {3 N_C \over 2}$, then
that model has a SUSY violating meta-stable ground state with SUSY
order parameter, $F \sim m_{ISS} \Lambda_{N_C} $.    $\Lambda_{N_C}$
is the confinement scale of the gauge theory, and the analysis is
under control for $m_{ISS} \ll \Lambda_{N_C}$.    The meta-stable
state also breaks a vector-like sub-group of the $SU(N_F)\times
U(1)$ flavor symmetry of the model, leading to a variety of
pseudo-goldstone bosons.    ISS also argued that a similar
meta-stable state existed in the model with $N_F = N_C$.   In terms
of the moduli space of the $m=0$ theory the vacuum was near the
point $M_i^j = P_i^A \tilde{P}_A^j / \Lambda_{N_C} = 0$, $B =
\tilde{B} = \Lambda_{N_C}$. $B$ and $\tilde{B}$ are the dimension
one interpolating fields for penta-baryons and anti-baryons made of
$5$ penta-quarks.   Note that these baryons are standard model
singlets and that the only flavor symmetry which is spontaneously
broken at this point in moduli space is the $U(1)$ penta-baryon
number (axial symmetries are explicitly broken by the mass term).

For $N_F \geq N_C + 1$ there is a well controlled effective field
theory of the meta-stable state for $m_{ISS}/\Lambda_{N_C} \ll 1$.
This is not true for $N_F = N_C$ .   Furthermore, as we will see
below, we will need to work in the region $m_{ISS} \geq \Lambda_5$
for phenomenological reasons.

The basic idea of this paper then, is to exploit the ISS meta-stable
vacuum of the $N_F = N_C = 5$ theory to construct an effective
theory for Cosmological SUSY Breaking (CSB).   The basic CSB input
is to fix the value of $m_{ISS}$ in terms of the CSB ansatz  for the
gravitino mass, $m_{3/2} \sim \lambda^{1/4}$ (where little $\lambda$
is the c.c.). This requires $m \sim {{\lambda^{1\over 4} m_P}\over
\Lambda_5 }$.   The factor of $\Lambda_5$ in the denominator of this
formula, anathema to an effective field theorist, can be
explained/excused in terms of the diagrams of \cite{susyhor}, which
mix up IR propagation through the bulk of space-time with UV
interactions with de Sitter horizon degrees of freedom.    One also
adds a constant to the superpotential to guarantee that the true
c.c. is in fact $\lambda$ in the low energy effective theory.   Both
of these terms break a discrete R symmetry of the $\lambda = 0$
theory.

The result is a lean and mean, stripped down version of the
Pentagon, suitable for rapid deployment to solve all\footnote{As is
conventional in communications from the Pentagon, we are here
indulging in a bit of hyperbole.} of the problems of the
supersymmetric standard model. It involves a single scalar field $S$
with discrete $R$ charge $2$. The only marginal/relevant couplings
of $S$ are encoded in a superpotential:
$$W_S = g_S S P_i^A P_A^j Y_j^i + g_{\mu} S H_u H_d  + g_T S^3.$$  $Y$ is the
unique traceless $SU(1,2,3)$ invariant matrix in the fundamental
representation of $SU(5)$.   Note that the $SU(1,2,3)$ standard
model gauge group is embedded in the obvious fashion into the vector
$SU(5)$ flavor group of the Pentagon.

In fact, we could consider a more general version of the model in
which we require only that the linear combination of penta-quark
bilinears to which $S$ couples, is linearly independent of the
combination to which $m_{ISS}$ couples.   The form we have written
{\it may} have a group theoretic justification at the unification
scale, which we will discuss below.

\section{Known knowns}

The full low energy Lagrangian of our model is
$${\cal L} = d^4 \theta\ [P^* e^{V} P + Q^* e^{V} Q + L^* e^{V} L +
( \bar{U})^* e^{V} \bar{U} + ( \bar{D})^* e^{V} \bar{D} + (
\bar{E})^* e^{V} \bar{E}]$$ $$ + [\int d^2 \theta\ ((\sum\ \tau_i\
W_{\alpha}^i )^2
 +  P_i^A \tilde{P}_A^j (m_{ISS} \delta_j^i + g_S S Y_j^i) + g_{\mu} S H_u
 H_d + g_T S^3$$ $$ + H_u Q_m (\lambda_u )^m_n \bar{U}^n + H_d Q_m (\lambda_d )^m_n
 \bar{D}^n + H_d L_m (\lambda_L)^m_n \bar{E}^n + {1\over M_U} L_m L_n \lambda^{mn}
 H_u^2 ) + h.c.] .$$ The scale of the neutrino seesaw operator is $M_U \sim
10^{14} - 10^{15}$ GeV.   We will take this parameter to be the
scale of all irrelevant corrections to the Lagrangian.

The gauge group of the model is $SU(5) \times SU(1,2,3)$, and the
sum over gauge kinetic terms sums over simple factors of this group.
When $m_{ISS}, g_S$ and the standard model gauge couplings are
turned off, the Lagrangian has an unbroken $SU_L (5) \times SU_R
(5)$ global symmetry, acting on the small Latin indices of the
penta-quarks $P$ and $\tilde{P}$.   The $SU(1,2,3)$ standard model
is embedded in the usual way in the vector (diagonal) $SU(5)$
subgroup of this chiral flavor group. Thus, if we use $SU(5)$
notation to summarize standard model quantum numbers then the $P$ is
in a $[5,5]$ under $SU(5) \times SU(1,2,3)$ while $\tilde{P}$ is in
a $[\bar{5}, \bar{5}]$.

The parameter $m_{ISS}$ is assumed to be induced by Cosmological
SUSY Breaking , CSB, as in \cite{susyhor}.  The c.c., $\lambda$ is a
tunable parameter\footnote{Actually it is discretely tunable. $\pi
(RM_P)^2 $ is the logarithm of an integer number of states.  $R$ is
the dS radius. In a more ambitious model based on holographic
cosmology, one gets a distribution of asymptotically dS universes
with different $\lambda$, and $\lambda$ can be anthropically
selected. However, this model has no other testable consequences
once the value of $\lambda$ is chosen, so there is no point in
discussing it here.}, and $m_{ISS}$ scales like $\lambda^{1/4}$ as
$\lambda \rightarrow 0$. In this limit, the low energy Lagrangian
has an $U(1)_R$ symmetry, which has no $SU_P (5)$ anomaly.   $P$ and
$\tilde{P}$ have R charge zero.  A discrete $Z_4$ subgroup of this
$U(1)$ is assumed to be an exact symmetry of the S-matrix when
$\lambda = 0$. $S$ has R charge 2. The R transformation properties
of the SSM chiral multiplets can be chosen so that the only
perturbative baryon or lepton number violating interaction of
dimension $\leq 5$ is the neutrino see-saw term,
$$\int d^2\theta\ {{(H_u L^m)(H_u L^n)S_{mn}}\over M_U} .$$ The
discrete R symmetry is also preserved by the 't Hooft interactions
induced by standard model instantons.

The R charge of $S$ does not permit the superpotential term $S^2$.
The linear term $S$ can be forbidden by a variety of strategies. The
simplest of these involves physics at the standard model unification
scale.   At that scale, we assume that the standard model is
unified, perhaps in a way that involves extra compact dimensions, in
a group containing the Georgi-Glashow group $SU(5)$. We further
assume that $S$ is the remnant of an $SU(5)$ adjoint, transforming
like the hypercharge generator of $SU(1,2,3)$. All other members of
this multiplet get mass at the unification scale. Finally, we assume
that no $SU(5)$ violating superpotential couplings of $S$ are
induced by the tree level breaking of $SU(5)$. SUSY
non-renormalization theorems then assure us that the terms
$$\int d^2\theta\ (a S + b S P_i^A \tilde{P}^i_A )$$ will not appear
in the low energy effective Lagrangian at the TeV scale.

The fact that $m_{ISS}$ produces SUSY breaking follows from the
neo-conservative revolution fomented by ISS.   These authors showed
rigorously that in SUSY QCD with $3 N_C / 2 \geq N_F \geq N_C + 1$ a
small mass term produces a meta-stable SUSY violating vacuum in
Poincare invariant quantum field theory.   For the indicated values
of $N_F$, a systematic small $m_{ISS}/\Lambda_{N_C}$ expansion of
the properties of this state could be established.   Giving a large
mass to {\it one} $SU(N_C)$ fundamental when $N_F = N_C + 1$, ISS
also argued that a similar meta-stable state existed for $N_F =
N_C$, though its properties were not under analytic control.   That
state had a non-vanishing expectation value for the penta-baryon
number violating operators
$$< \epsilon^{A_1 \ldots A_{N_C}} P_{A_1}^{i_1} \ldots
P_{A_{N_C}}^{i_{N_C}}> = \Lambda_{N_C}^{N_C} \epsilon^{i_1 \ldots
i_{N_C}},
$$
$$< \epsilon_{A_1 \ldots A_{N_C}} \tilde{P}^{A_1}_{i_1} \ldots
\tilde{P}^{A_{N_C}}_{i_{N_C}}> = \Lambda_{N_C}^{N_C} \epsilon_{i_1
\ldots i_{N_C}} . $$  By contrast, the meson operators $P^A_i
\tilde{P}_A^j $ have vanishing expectation value in this state.

This model is merged with the hypothesis of CSB by making two
assumptions.   First the parameter $m$ is determined so that the
gravitino mass at the meta-stable SUSY violating vacuum is given by
the CSB formula
$$m_{3/2} = \gamma \lambda^{1/4} (M_p / m_P), $$ where $M_P = \sqrt{8\pi}
m_P$ is the Planck mass, and $\gamma$ is an unknown constant,
expected to be of order one. This means that
$$m_{ISS} \sim \gamma {{\lambda^{1/4} M_P}\over \Lambda_5} .$$   Note that
this is a term in the Lagrangian at a scale $\gg \Lambda_5$ where
the Pentagon interactions are weak.   From the point of view of
standard effective field theory, it is extremely peculiar to have
the IR scale $\Lambda_5$ appear in the denominator of this
parameter.   However, the diagrams contributing to the argument for
the CSB formula that I presented in \cite{susyhor}, combine infrared
propagation in a single horizon volume, with UV dynamics in the
vicinity of the horizon.   It is plausible that they contain such
inverse IR scales.

The second input from CSB is that we add a constant $W_0$ to the
superpotential to tune the c.c. at the meta-stable minimum to
$\lambda$.   Again, there is no reason to do this in effective field
theory, though in this case it would be the strategy of any
effective field theorist who wanted the meta-stable vacuum to be of
phenomenological relevance.   The logic for adding $W_0$ in CSB is
different.   $\lambda$ is viewed as a high energy input parameter
determined by the number of quantum states in the asymptotic de
Sitter space to which the system is converging.   It cannot be
renormalized, and the constants in effective field theory must be
tuned to reproduce its input value.

It should be emphasized that from the point of view of low energy
effective field theory, the model is defined without reference to
CSB.   Thus, the effective field theorist can simply take the mass
term $m_{ISS}$ to be a parameter of unknown provenance, or imagine
that it arises as a consequence of the dynamics of another gauge
group, not included in the Pentagon model.   Our insistence on the
origin of $m_{ISS}$ in CSB is a strong constraint, because it bounds
the scale of SUSY breaking in the model.   We will see that this
makes it more difficult to find a working version of the model.

One other place where CSB plays a role in our considerations is our
decision to tune the c.c. (almost) to zero near a particular SUSY
violating minimum of the effective potential of the flat space field
theory. It would not be consistent with the hypothesis of CSB to
perform this tuning at the SUSic minimum.    We will in fact have to
choose between two SUSY violating minima of the potential on
phenomenological grounds.   It is important to note that once we
have done this tuning, the supersymmetric states no longer have
anything to do with the theory.   Our meta-stable SUSY violating
world can tunnel to the negative c.c. region of the potential, but
the resulting bubble is a low entropy, highly non-supersymmetric,
Big Crunch geometry.   These tunneling amplitudes are too small to
be of any interest.

Before sailing for murkier waters, in which we will have to swim in
order to get interesting phenomenological consequences of this
model, I will list the approximate symmetries of the low energy
Lagrangian. The pure $N_F = N_C = 5$ model has a $U(5) \times U(5)$
symmetry. The axial $U(1)$ is an R symmetry under which $P$ and
$\tilde{P}$ have R charge 0.   This is also a symmetry of the Yukawa
couplings of the $S$ field, if $S$ has R charge $2$ and the up and
down Higgs fields have opposite R charge.   On the other hand the
mass term $m_{ISS}$ breaks $U(1)_R$ and is required to be fairly
large for phenomenological reasons.   This term also breaks $SU(5)
\times SU(5)$ to its diagonal subgroup.   The Yukawa coupling $g_S$
and the standard model gauge couplings break the diagonal $SU(5)$
down to $SU(1,2,3)$ of the standard model.    The meta-stable ISS
vacuum spontaneously breaks the remaining $U(1)$ Penta-baryon
symmetry. The Goldstone boson is the field defined by
$$<B> = \Lambda_5 e^{i b/\Lambda_5} = <\tilde{B}>^* ,$$
in the ISS vacuum.  We call this the {\it penton} field, though a
catchier name might be found, if one were willing to put in the
effort to think about it.

In addition to these continuous low energy symmetries, the Pentagon
has an anomaly free $Z_5$ subgroup of the axial symmetry which gives
$P$ and $\tilde{P}$ charge one.   The mass term $m_{ISS}$ breaks
this symmetry as well as the $U(1)_R$ but preserves a diagonal
$Z_5^R$ subgroup.   This is inconsistent with the tenets of CSB,
according to which the c.c. breaks all R symmetries.  It would also
prevent us from generating gaugino masses. The couplings $g_S$ and
$g_T$ violate this $Z_5^R$ if they are both non-zero.   These
couplings must therefore be large at low energy.

The usual chiral Lagrangian predictions for a Goldstone boson relate
the emission amplitude for a single Goldstone boson in a transition
between two final states to the change in the spontaneously broken
quantum number in the transition.    This would predict zero
coupling to ordinary quarks and leptons since they do not carry
penta-baryon number.   However, there are dimension $5$ operators
 $${c_B \over \Lambda_5} \partial_{\mu} b J^{\mu}_B + {c_L \over
 \Lambda_5} \partial_{\mu} b J^{\mu}_L , $$ which are allowed by the
 symmetries of the low energy Lagrangian.   These are in fact
 generated by the diagrams of Figure 1, involving standard model
 couplings at the scale $\Lambda_5$.   The nominal estimates for the
 coefficients are
 $$c_B \sim \alpha_3^2 (\Lambda_5 ), \ \ \ c_L \sim \alpha_2^2
 (\Lambda_5 ) ,$$ where these are running gauge couplings, at the
 indicated scale.

\begin{figure}[ht!]
\unitlength.5mm \SetScale{1.418}
\begin{center}
\begin{picture}(130,45)(80,-10)
\DashLine(30,0)(60,0){2}
\ArrowArcn(85,0)(25,0,180)\ArrowArc(85,0)(25,0,180)
\Gluon(105,15)(179,15){6}{4} \Gluon(105,-15)(179,-15){-6}{4}
\ArrowArcn(199,0)(25,0,180)\ArrowArc(199,0)(25,0,180)
\DashLine(226,0)(256,0){4} \Text(85,23)[t]{$P$}
\Text(85,-23)[b]{$\tilde{P}$} \Text(199,-23)[b]{$\tilde{Q}$}
\Text(199,23)[t]{$Q$}
\Text(45,0)[b]{$\partial_{\mu} b$} \Text(241,0)[b]{$J_{\mu}^F$}
\end{picture}
\end{center}
\vspace{10mm} \caption{Diagrams leading to penton interaction with
standard model currents. Gauge bosons are any charge and color
neutral pair in $SU(1,2,3)$. The RHS loop could contain leptons.}
\label{fig:vertex}
\end{figure}
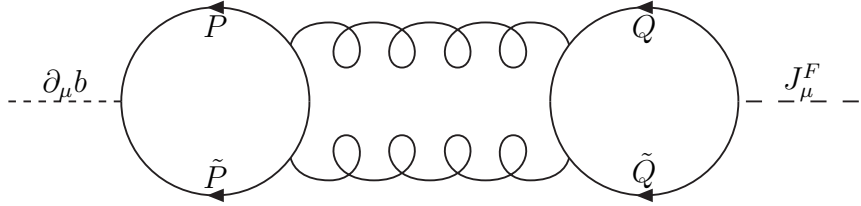

 These terms in the action are total derivatives if we neglect the
 violation of baryon and lepton number by electroweak instantons,
 but they will be important in the early universe if there are
 temperatures as high as the electroweak sphaleron mass.

 Under laboratory conditions, the dominant dimension 5 coupling of the penton to
 normal matter comes from similar dimension five couplings to
 non-conserved neutral hadronic currents like beauty, charm, and strangeness.
 They are suppressed by two powers of $\alpha_3 (\Lambda_5 )$ and are proportional to
 quark mass differences in units of the QCD scale (for light quarks).   They are
 of the form
 $$\alpha_3^2 {m_F \over \Lambda_{QCD}}{{\partial_{\mu} b}\over \Lambda_5} J^{\mu}_F , $$ where
 $J^{\mu}_F$ is a flavor current.   Note that analogous couplings to axial currents
 are suppressed by further powers of electroweak gauge couplings,
 because the $SU(5) \times SU(3)$ gauge theory is invariant under parity. Thus pentons will be
 predominantly emitted in charged current weak decays, since the
 amplitudes are proportional to the divergence of the corresponding
 flavor current.   As we will see in the section on dark matter, pentons
 are very light, $.1 - 1$ eV, and will escape from the detector as missing energy.

Models with very low energy SUSY breaking cannot contain the usual
SUSic candidates for dark matter.   Even if R parity is conserved,
the LSP is the gravitino, which is relativistic at the usual scale
of matter domination.    It is also relatively strongly coupled and
so the NLSP is not cosmologically meta-stable.   In previous
attempts to construct a phenomenology based on CSB, I suggested that
baryons of the new strong interactions (penta-baryons in the
Pentagon model) would be the dark matter.   This is not possible if
penta-baryon number is spontaneously broken.    Instead I will
suggest that the dark matter is a condensate of pentons.   I will
show that this is reasonable if the early universe produces a
sufficiently large penta-baryon asymmetry.

\section{\bf Known unknowns}

We will accept the hypothesis of ISS that the Pentagon model has a
meta-stable SUSY violating state with flat space vacuum energy of
order $m_{ISS}^2 \Lambda_5^2$.   ISS characterize this as a state
which has vanishing expectation value for penta-meson operators.
However, in the presence of $m_{ISS}$ and the couplings $g_S$ and
$g_T$ there is no symmetry which prevents the combinations $P_i^A
\tilde{P}^j_A (\delta^i_j, Y^i_j)$ from getting non-zero VEVs. Thus,
we expect these bilinears to have VEVs of order $K \Lambda_5^2$ at
the SUSY violating minimum, where $K$ involves powers of $g_{S,T}$
if these are small.  We will tune the parameter $W_0$ in the SUGRA
formula for the effective potential, so that the c.c. at this SUSY
violating minimum is of order the observed c.c., $\lambda$.

Actually we must resolve one further ambiguity in choosing the SUSY
violating vacuum.   When $m_{ISS} = 0$ there are two solutions of
the F and D term constraints for the Higgs fields and the singlet
$S$:

$$h_{u,d} = 0, \ \ \ g_S p_i^A \tilde{p}^j_A  Y^i_j = - 3 g_T s^2
,$$ and
$$s = 0 , \ \ \ g_S p_i^A \tilde{p}^j_A  Y^i_j = - g_{\mu} h_u h_d
,$$ where lower case letter represent the scalar components of
chiral superfields.   The Higgs VEVs in the second equation are
oriented so that electromagnetism is unbroken and ${\rm tan} \beta$,
the ratio of Higgs VEVs is one.

In the presence of the SUSY breaking parameter $m_{ISS}$ there will
be a similar ambiguity in the choice of VEVS at the SUSY violating
vacuum.   However, $s= 0$ is no longer a stationary point of the
effective action since there are no unbroken symmetries which
preserve it.   On the other hand, the first minimum will still have
$h_{u,d} = 0$.    We will choose to tune the c.c. to $\lambda$ at
the stationary point where $SU(2) \times U(1)$ is broken .   The
other SUSY violating stationary point of the flat space potential
may then have either positive or negative c.c., while the erstwhile
SUSic states will have negative c.c.  .   None of the other
stationary points of the potential represent long lived states of
the universe once gravity is taken into account.

The Pentagon model thus has a stationary point of its effective
potential with spontaneously broken SUSY ( $F \sim m_{ISS}
\Lambda_5$), and $SU(2) \times U(1) \rightarrow U(1)_{EM} $ with
$|h_u| \sim |h_d| \sim \Lambda_5$, which we will take to represent
the real world.

SUSY breaking is communicated to the standard model by two distinct
mechanisms.   Since $s,h_u, h_d$ are all non-zero, the Higgs
superfields have F terms which will contribute tree level masses to
squarks and sleptons.   In addition there are more or less
conventional gauge mediated contributions.    The latter are the
dominant contributions for gaugino masses as well as the masses of
squarks and sleptons, apart from the top squark. Gaugino masses are
estimated from one loop standard model diagrams with pentaquark
superfields in the loop, and arbitrary numbers of penta-gauge
bosons. If we compare to conventional gauge mediated scenarios,
these diagrams are enhanced by a factor of $5 \sum Y^2 \sim 16.7 $
(where the sum is over the weak hypercharges in the 5 representation
of $SU(5)_{GUT}$). As we will see, this means that the gaugino to
squark or slepton mass ratios are larger by a factor $\sim 4$ than
they are in conventional gauge mediated models.

The corresponding two loop contribution to {\it e.g.} the right
handed charged slepton squared mass is enhanced by the same factor.
Thus
$$m_{\tilde{e}_R} \sim 4 {\alpha_1 \over \pi} {F\over\Lambda_5 } .$$
In the standard model ${\alpha_1 \over \pi} \sim {1\over 250} $ so
we need $F/\Lambda_5 > 6.25 $ TeV in order to satisfy the
experimental bounds on this mass.   The CSB prediction for $F$ is of
order $10$ TeV$^2$ so this implies $\Lambda_5 \sim 1.6$ TeV.   The
ISS mass parameter then satisfies
$${{m_{ISS}}\over \Lambda_5} \sim 4 .$$
It is expected\footnote{A proof of this fact has not yet been
found.} that when $m_{ISS} \gg \Lambda_5$, the meta-stable state of
ISS disappears.  One would expect to be able to integrate out the
penta-quarks at a scale where the Pentagon gauge coupling was weak,
leaving over the pure $SU(5)$ gauge theory and the supersymmetric
standard model, coupled only by irrelevant operators.   This theory
does not have any meta-stable SUSY violating states. The ISS
analysis itself was carried out in the limit where $m_{ISS} \ll
\Lambda_5$.

The phenomenologically preferred value for $m_{ISS}/\Lambda_5$ is
not in the perturbative regime.   For example, a corresponding ratio
in QCD would correspond to quark masses of order $600$ MeV. Thus, it
is not implausible that the meta-stable state exists in the
phenomenologically required region. Note also that the additional
hypercharged states in the Pentagon model might make the coupling
$\alpha_1$, which appears in the estimate for the slepton mass,
slightly larger, and consequently loosen the bound on $F/\Lambda_5$.
Nonetheless, these considerations suggest that the lightest charged
sleptons in the Pentagon model of SUSY breaking cannot be
significantly heavier than the current experimental lower limits.
Thus, we expect to see these sleptons produced at LHC. Since the LSP
is the light gravitino, with couplings of order $1/F$, slepton pair
production will result in spectacular final states with two hard
leptons, other hard particles, and missing energy, a classic signal
for low energy SUSY breaking\cite{classic}.  It is however likely
that the NLSP in this model is not a gaugino, because of the extra
factor of $4$ in gaugino masses.   The other hard particles in the
final state depend on the nature of the NLSP, which we cannot
determine at this time.

It is worth pointing out that these estimates of the scale of SUSY
breaking give a gravitino mass of order $5 \times 10^{-3}$ eV. Such
gravitinos are perfectly consistent with Big Bang Nucleosynthesis,
in sharp contrast to conventional gauge mediated models.   They are
light enough, and their longitudinal components strongly coupled
enough, that one might imagine finding them in experiments probing
for short distance modifications of gravity.  It has also been
suggested that they might be found at the LHC\cite{zwirner}.

Another difference between the Pentagon model and conventional gauge
mediated models is that the $SU(2)\times U(1)$ violating top squark
mass generated by the $F$ term of $H_u$ is comparable to or larger
than the gauge mediated mass.   This is because our model
effectively generates a sizable effective $\mu$ term, from the VEV
of $S$.

  Unfortunately, the
strong coupling physics at the scale $\Lambda_5$ prevents us from
making very precise statements about the spectrum of superpartner
masses.    In particular, there are three potentially worrisome
tuning problems that I have not had the calculational skill to
address. First, our estimates of gaugino and right handed slepton
masses depended on the assumption that the couplings $g_S$ and $g_T$
were strong enough that there is no further loop suppression of
gaugino masses (recall that these couplings broke the discrete R
symmetry left over by $m_{ISS}$).  In particular one may worry that
$g_T$ is not asymptotically free (it would appear to be renormalized
only by wave function renormalization of the gauge invariant $S$
field) so that a large value at low energy may lead to a Landau pole
well below the unification scale.

The second potential tuning is the ratio of the electroweak scale,
$250$ GeV to $\Lambda_5$.   For $g_S$ of order $1$, this is of order
$1/6$.    It scales like $g_S$ for small $g_S$, since the VEV of the
bilinear $PY\tilde{P}$ is of order $g_S$, but small $g_S$ would
alter our estimates for gaugino and charged slepton masses in an
unpleasant fashion. Finally, one may worry about the \lq\lq little
hierarchy problem".   Precision electroweak measurements seem to
prefer a Higgs mass below $200$ GeV, and this may also be a little
tuned in the current model.   It is hard to tell whether one should
take factors of $6 - 10$ seriously in a model where it is so hard to
make precise calculations.

One thing that appears safe is direct interference of the Pentagon
degrees of freedom with precision electroweak measurements.   These
would primarily affect the Peskin-Takeuchi S-parameter, but with our
estimates of $\Lambda_5$ and $m_{ISS}$ the effects seem to be
small\footnote{I would like to thank H. Haber for conversations
about this point.}. These same estimates suggest that the expected
rich \lq\lq penta-hadron" spectrum may be beyond the discovery reach
of the LHC. Scaling up from QCD we might expect penta-mesons in the
$6 - 10$ TeV range. Penta-baryons will probably be unstable to
decays into penta-mesons and pentons, with life-times of order
$\Lambda_5^{-1}$. The penton itself is the only light remnant of the
penta-hadron spectrum.   As discussed above, it should be produced
in association with ordinary charge changing weak decays and can be
searched for in low energy experiments, rather than the LHC.

\section{Baryon number, lepton number, and flavor}
\subsection{B and L violating operators of dimension $4$ and $5$}

A central element in CSB is the discrete R symmetry which guarantees
Poincare invariance in the the limiting model. This can be put to
other uses.   In \cite{susycosmophenoIII} I showed that it can
eliminate all unwanted dimension $4$ and $5$ baryon and lepton
number violating operators in the supersymmetric standard model.
This is sufficient to account for experimental bounds on baryon and
lepton number violating processes.  The interaction $\int d^2
\theta\ H_u^2 L^2$, should not be forbidden by R.  I will adopt the
philosophy of a previous paper and insist that the texture of quark
and lepton Yukawa couplings, as well as neutrino masses, are
determined by physics at the unification scale.

We will choose the R charge of SSM fields to be independent of quark
and lepton flavor, and denote it by the name of the corresponding
field. All R charges are to be understood modulo $N$, where $Z_N$ is
the R symmetry group. In the remodeled Pentagon model, we must
choose $N = 4$ in order to accommodate the $g_T S^3$ term in the
superpotential.   We will also impose anomaly freedom for the
discrete R symmetry.  That is, the 't Hooft interactions generated
by all instantons should be invariant.  This leads to the three
constraints (all equations in this section are equalities mod 4):
\eqn{anom5}{5 (P + \tilde{P} ) = 0}\eqn{anom3}{6 Q + 3 \bar{U} + 3
\bar{D} + 5(P + \tilde{P}) = 0}\eqn{anom2}{H_u + H_d + 9Q + 3L + 5(P
+ \tilde{P}) = 0 }  In writing these equations, we have taken into
account the gaugino charges, and dropped terms that are explicitly
zero mod 4.   Using the first anomaly equation and dropping more
terms which vanish mod 4, the second two equations can be simplified
to: \eqn{anom3a}{6Q + 3(\bar{U} + \bar{D}) = 0 ,} \eqn{anom2a}{3Q +
L = 0 .}

The condition that the standard Yukawa couplings are allowed by R
symmetry is

\eqn{yukallow}{L + H_d + \bar{E} = Q + H_d + \bar{D} = Q + H_u +
\bar{U} = 2.} The coupling $ S H_u H_d$ requires \eqn{sallow}{H_u +
H_d = 0.}

Note that these conditions forbid the standard $\mu$ term $\int d^2
\theta\ H_u H_d$, .  We will also impose $2L + 2H_u = 2$ to allow
the dimension $5$ F term which can generate neutrino masses. The
renormalizable dynamics of the Pentagon gauge theory preserves all
flavor symmetries of the standard model. This forbids the generation
of the neutrino seesaw term with coefficient ${1 \over \Lambda_5}$.
As emphasized in \cite{susycosmophenoIII}, we imagine the neutrino
seesaw term, and the texture of the quark and lepton mass matrices,
to be determined by physics at the scale $M_U$, probably via a
Froggat Nielsen mechanism.

Dimension $4$ baryon and lepton number violating operators in the
superpotential will be forbidden in the limiting model by the
inequalities

\eqn{nobla}{2 L + \bar{E} \neq 2} \eqn{noblb}{2 \bar{D} + \bar{U}
\neq 2, } \eqn{noble}{L + Q + \bar{E} \neq 2 .} Absence of
dimension $5$ baryon number violating operators requires
\eqn{noblc}{3Q + L \neq 2}\eqn{nobld}{3Q + H_d \neq
2}\eqn{noblf}{\bar{E} + 2 \bar{U} + \bar{D} \neq 2 ,}

The condition that there be no baryon number violating dimension
$5$ D-terms is that none of $ Q + \bar{U} - L ; $ or $U + E - D$,
vanishes.

We can solve for all of the R charges in terms of $L$ and $H_d$:

$$ Q = - 3L$$ $$\bar{E} = 2 - L - H_d$$ $$\bar{D} = 2 + 3L - H_d$$
$$\bar{U} = 2 + 3L + H_d.$$ In addition we have the relation
$$ 2L = 2 H_d + 2 .$$  The inequalities which forbid dangerous
operators are all satisfied if and only if $L$ is odd and $H_d$ is
even.   Any choice satisfies the last constraint, so we have four
solutions $L = \pm 1$, $H_d = 0,2$ .

\subsection{Flavor and CP}

The Pentagon shares with generic gauge mediated models the property
that the only terms in the low energy Lagrangian that are not
invariant under the $SU(3)_Q \times SU(3)_{\bar{U}} \times
SU(3)_{\bar{D}}$ flavor group of the standard model, are the quark
and lepton Yukawa couplings, and the neutrino seesaw term. As a
consequence it has a GIM mechanism, and flavor changing neutral
currents are suppressed below experimental upper bounds.  Similarly,
lepton flavor changing processes like $\mu \rightarrow e + \gamma$
are within experimental limits.   Quark and lepton flavor changing
processes, in addition to those induced by the neutrino seesaw term,
will come from dimension $6$ operators. We can assume that they are
scaled by the same operator $M_U \sim 10^{15} $ GeV as the neutrino
seesaw.   Note that the restriction to dimension $6$ operators is
non-trivial and depends on the fact that dimension $5$
superpotential terms like $Q^2 \bar{U} \bar{D} $ (with various
flavor combinations) are forbidden by the $Z_4$ R symmetry, although
they conserve baryon and lepton number.  Flavor violation comes
predominantly through SSM loop graphs.

The remodeled Pentagon thus solves most of the problems of generic
SUSic models.  However, it does have CP violating phases in addition
to the usual CKM parameter.    To see this, perform the following
sequence of transformations.
\begin{itemize}

\item A $U_A (1)$ transformation on quarks, to eliminate the QCD $\theta$ angle, $\theta_3$.

\item A similar $U_AP (5)$ transformation on penta-quarks to eliminate $\theta_5$.

\item An anomaly free $U_R (1)$ transformation, to eliminate ${\rm arg} m_{ISS}$.

\item A common phase rotation on $H_{u,d}$ to eliminate ${\rm arg det} \lambda_u \lambda_d$.

\item A phase rotation of $S$ to eliminate ${\rm arg} g_{\mu}$.

\end{itemize}

We are left with the phases of $g_{S,T}$ (as well as phases in the
neutrino see-saw term and CKM matrix) as physical CP violating
parameters. When we integrate out scales $\gg \Lambda_{QCD}$, these
phases will infect the determinant of the renormalized quark mass
matrix and are likely to give rise to an electric dipole moment for
the neutron which is incompatible with experimental bounds.  The
upside of this result is that the potentially troublesome axion,
which roamed the halls of the old Pentagon, no longer exists.

Counting one of these two phases as a stand in for $\theta_3$ we see
that the Pentagon model has six new physical parameters, the
absolute values of $m_{ISS}, \Lambda_5 , g_{S,T,\mu}$ and one
combination of the phases of $g_{S,T}$ in addition to the parameters
in the standard model Lagrangian.   Of these, $|m_{ISS}|$ is roughly
determined by the rules of CSB.   The three Yukawa couplings of the
$S$ field are required to be reasonably large.   We will discuss the
consequences of this assumption in the section on unification of
couplings.

\section{\bf Dark matter}

CSB, like any model with a maximum SUSY breaking scale of order $<
100$ TeV, does not have a cosmologically stable massive LSP.  Even
if R-parity is preserved, the LSP is the gravitino, and its
longitudinal components are so strongly coupled that the NLSP will
decay to it rapidly, probably in typical particle detectors, and
certainly with non-cosmological lifetimes. In previous discussions
of CSB, I have suggested baryons of the new strong gauge group as
dark matter candidates and in \cite{darkg} we showed that this was a
viable option if a sufficiently large penta-baryon asymmetry is
generated in the early universe. The meta-stable ISS SUSY breaking
state also breaks penta-baryon number spontaneously, so
penta-baryons are no longer cosmologically stable.

Instead, I want to show that with a sufficiently large penta-baryon
number asymmetry, the penton can be the dark matter.   Assume a
penta-baryon asymmetry to entropy ratio $\epsilon$ is generated in
the early universe at or after inflationary reheating, and that
there is no significant entropy production thereafter.    The
universe is radiation dominated and the penta-baryon density at
temperature $T$ will be

$$n_{PB} = \epsilon G T^3,$$ where $G$ counts the effective number
of massless degrees of freedom in the plasma.   Once $T$ drops below
$\Lambda_5$ the penta-baryon density can be written in terms of the
penton field $b$
$$n_{PB} = \Lambda_5 \dot{b}, $$ so we get the equation of motion
$$\dot{b} = \epsilon G {{T^3}\over \Lambda_5} .$$

So far we have not taken explicit penta-baryon number violation into
account.  The leading gauge invariant supersymmetric penta-baryon
number violating interactions are the dimension six F terms ${1\over
M_U^2 } \int d^2 \theta\ (a P^5 + b \tilde{P}^5 ) $.   We will take
the mass scale in these interactions to be the same order of
magnitude, $\sim 10^{15}$ GeV, as that which enters the dimension
$5$ operator that leads to neutrino masses and mixings. Below the
scale $\Lambda_5$ these interactions will give rise to a potential
for $b$ of order
$${{\Lambda_5^6} \over M_U^2} V(b/\Lambda_5 ), $$ where $V$ is a
periodic function.  This gives a penton mass of order
${{\Lambda_5^2}\over M_U} \sim 2 eV$.

The potential will begin to affect the cosmological evolution of the
penton when the kinetic energy generated by the asymmetry is of
order the potential.   This happens at a temperature $T^*$ given by
 $$\epsilon^2 G^2 (T^*)^6 = {{\Lambda_5^8 }\over M_U^2}.$$   Unless
 $\epsilon G < {\Lambda_5 \over M_U} \sim 10^{-12} ,$ this
 temperature is below $\Lambda_5$ and so our description of the
 effects of the explicit symmetry breaking is valid.   We will see
 that $\epsilon$ has to be quite large if we want the penton to be
 dark matter.

 Indeed, below the scale $T^*$ the penton density will grow relative
 to the radiation energy density by a factor ${T^* \over T}$.
 $${\rho_p \over \rho_{\gamma}} = \epsilon^2 G ({T^* \over
 \Lambda_5})^2 {T^* \over T}.$$  Note that
 $$({T^* \over \Lambda_5})^3 = {\Lambda_5 \over {\epsilon G M_U}},$$
 so that
 $${\rho_p \over \rho_{\gamma}} = \epsilon {\Lambda_5 \over { M_U}}{\Lambda_5 \over
 T}.$$   Matter radiation equality occurs at $T \sim 10 $ eV in the
 real world, and we can achieve this if $\epsilon \sim 5$.   Values
 of $\epsilon$ this large can probably only be achieved via coherent
 classical processes analogous to Affleck-Dine baryogenesis, but are certainly not implausible
 in that context.

 We remarked in the second section of this paper that there is a
 coupling between the penta-baryon and baryon number currents
 induced by QCD interactions above the scale $\Lambda_5$. In the
 presence of the asymmetry we have postulated here, this gives rise
 to a time dependent chemical potential for baryon number in the
 early universe.   It would be remarkably interesting if, when
 combined with ordinary electro-weak sphaleron processes, this could
 give rise to the observed baryon asymmetry, thus tying together the
 baryon asymmetry and dark matter densities of the universe.  This
 would be a form of {\it spontaneous baryogenesis}, as first
 envisaged by Cohen and Kaplan\cite{ck} .    We hope to report
 on this interesting possibility in the near
 future\footnote{Preliminary analysis and further work on this
 problem have been done in collaboration with S.Echols and
 J.Jones\cite{tbej}.}.

\section{Coupling constant unification}

The extra matter in the Pentagon model consists of the $SU_P(5)$
gauge multiplet, the $SU(1,2,3)$ singlet $S$ and the the
penta-quarks, which are in complete multiplets of the $SU(5)$ GUT
group.   One loop gauge coupling unification will not be affected by
these new states, but the value of the unified gauge coupling is
considerably enhanced. Indeed, the beta function for the $SU(2)$
coupling above the scale $m_{ISS}$ is
$${{d\alpha_2^{-1}}\over dt} = - {6\over 2\pi} ,$$
which gives a value of $\alpha_2^{-1}$ slightly less than $8$ at the
unification scale.   The Landau pole in this one loop running
coupling comes at $\sim 7 \times 10^{19}$ GeV, so we seem to be just
within the perturbative regime  .   Two loop calculations make the
unified coupling even larger and one may be skeptical of the
perturbative expansion.   Nonetheless, it appears reasonable to
claim that this model predicts coupling unification.   The large
size of the unified coupling suggests that dimension six operators
may give proton decay within range of future experiments.

A more troubling problem is posed by the Yukawa couplings $g_{S,T}$.
These are required to be large at the scale $\Lambda_5$ in order to
provide sufficiently large gaugino masses. While it is possible that
the physical (as opposed to holomorphic) $g_S$ is asymptotically
free (since it has a large negative term in its one loop $\beta$
function coming from Pentagon gauge interactions), this is not true
for $g_T$.   It is thus likely that the value of $g_T$ that we need
for acceptable gaugino masses will lead to a Landau pole below the
unification scale.   Further investigation is required to determine
whether the prediction of coupling unification can be salvaged.

\section{Anthropic considerations}

Viewed as an effective field theory, the Pentagon model has a
coincidence of scales, $m_{ISS} \sim 4 \Lambda_5$, which is forced
on us by phenomenology.   One might try to explain this coincidence
by anthropic arguments.   I will discuss only the version of this
argument that follows from the principles of CSB, in which only the
parameter $m_{ISS}$, which is determined in terms of the c.c., is
allowed to vary.   The whole structure of the Pentagon model,
determined as it is by discrete $R$ symmetries, does not fit in well
with the String Landscape\cite{Dine}, in which we would have to
assume that all parameters are anthropically scanned.   Within the
context of CSB, Weinberg's anthropic bound on the c.c takes on its
full force, and one does not have to worry about varying other
parameters.

However, if the c.c. is an input parameter, governing the number of
states in the quantum theory, it is no longer safe to assume that
the probability distribution determining it is flat near $\Lambda =
0$. For example, a flat distribution in the number of states
corresponds to a strong preference for very small $\Lambda$.
Weinberg's argument that we observe a typical value for the c.c.
that allows galaxies to exist is no longer so obvious. A
meta-physical model, which introduces an {\it a priori} preference
for large $\Lambda$ \cite{holocosm} could solve this problem.

It is however interesting that the qualitative low energy physics of
our model changes drastically as soon as $\sqrt{\lambda^{1/4} M_P}
\sim 100$ GeV rather than $\sim 3 $ TeV.  This corresponds to
reducing $m_{ISS}$ by a factor of $30$ so that $m_{ISS}/\Lambda_5
\sim .1$ .   When this dimensionless parameter is small the low
energy theory has a large set of degrees of freedom charged under
the standard model and the QCD coupling does not become
asymptotically free until a scale of order $ 100 $ GeV .   We should
imagine that its short distance value is fixed, so that $\alpha_3 (
100 {\rm GeV}) $ is much smaller than its experimental value.   With
a very crude estimate, we find that this reduces the QCD scale by a
factor of $200$.   Note that the scale $\Lambda_5$, and thus the
scale of electroweak interactions is unchanged, while the value of
the electroweak couplings $g_{1,2}$ is slightly reduced.   Quark and
lepton masses are not changed significantly.  It is clear that these
changes will have dramatic effects on nuclear physics and stellar
evolution, and bring the scales of atomic and nuclear physics closer
together.   Such changes make life of our type impossible.   More
work would be necessary to determine precisely what the anthropic
lower bound on $\lambda$ is in this framework, but it is clear that
it is within a few orders of magnitude of the observed value.

One might also ask whether an improved {\it upper} bound on
$\lambda$ would follow from a similar argument.   This seems
unlikely, since the physics of the Pentagon model depends only on
the fourth root of $\lambda$.   However, there is one way in which a
tight upper bound on $\lambda$ might arise from the Pentagon.
Suppose that the phenomenologically required value of
$m_{ISS}/\Lambda_5 $, which appears to be $\sim 4$, were close to
the value at which the meta-stable ISS state of the Pentagon model
disappears.  If {\it e.g.} the critical value of this ratio were
$10$, then increasing $\lambda$ by a factor of $(2.5)^4 \sim 40$
would completely change the low energy world.

Indeed, whether or not it gives us a strong anthropic bound on
$\lambda$, the existence of a critical value of $r \equiv
m_{ISS}/\Lambda_5$ raises an interesting conundrum for CSB. The
basic idea of CSB is that the finite $\lambda$ theory is described
by a finite number of fermionic oscillators, which represent
quantized pixels on the cosmological horizon of dS space\cite{tbdS}
.   This finite system has an approximate $S$ matrix, which
converges to the $S$ matrix of a super-Poincare invariant theory of
quantum gravity in the $\lambda \rightarrow 0$ limit.   Low energy
matrix elements of this approximate $S$ matrix are supposed to be
calculable from the Pentagon model for sufficiently small $\lambda$.
If there is a critical value of $r$ above which the Pentagon model
does not break SUSY, then, above this value of r, the low energy
Pentagon model is not a good approximation to the underlying theory
of pixels.   That model certainly does not have exact SUSY, and
certainly does have a finite number of states.   On the other hand,
since the critical value of $\lambda$ is $\ll m_P$, one would
imagine that there is a valid low energy Lagrangian for the system
even above the critical value.   What is it? and how can we
understand the transition between the two descriptions in low energy
terms?

\section{\bf Conclusions}

The remodeled Pentagon is a much more robust structure than its
predecessor.  The existence of a meta-stable SUSY violating state is
on fairly firm footing and the complicated singlet sector of the
previous model has been replaced by a single field.   This
eliminated the troublesome low scale axion.   Elimination of an
approximate residual $Z_5$ R symmetry seems to restrict the
fundamental R symmetry group of the model to be $Z_4$.  The model
has no problems with FCNC or unwanted baryon and lepton number
violating interactions of dimension less than six.   Its unification
scale couplings are large enough that dimension six proton decay
might occur at observable rates.   Precision electroweak
measurements also seem to pose no problems for the Pentagon, since
the particles in the new sector (apart from the neutral penton) are
in the 6-10 TeV range.

The Pentagon model predicts relatively light charged sleptons, which
will be produced at LHC and have the spectacular decay signals of
sleptons in all SUSY breaking scenarios with a very light gravitino.
The final state includes opposite sign hard leptons plus missing
transverse energy plus $X$.   It is possible, but unlikely since
gauginos appear to be heavy in our models, that $X$ is the two
photon signal of standard gauge mediation.  The nature of $X$ is
determined primarily by the identity of the NLSP in the Pentagon
model and our ability to calculate the details of the sparticle
spectrum in the Pentagon model is limited.   One should also note
that, as a consequence of the low SUSY breaking scale, decays to
gravitinos will be prompt in our model and there will be no
displaced vertices.

 The Pentagon gives a pattern of SUSY breaking similar to but distinct
from conventional gauge mediated models. The typical squark or
slepton mass ratio is enhanced by a factor $\sim 4$ (but this only
effects the determination of the scales of the model in terms of
experimental quantities). However, the typical gaugino to squark or
slepton mass ratio is also enhanced by a factor $\sim 4$. Top
squarks have relatively large $SU(2)\times U(1)$ breaking masses
coming from the F terms of the Higgs fields, which may be comparable
to or larger than their gauge mediated masses. Electroweak symmetry
breaking occurs naturally, with the right pattern and ${\rm tan}
\beta \sim 1$.

Dark matter consists of the penton, a scalar pseudo-Goldstone boson
of penta-baryon number.   This requires a penta-baryon asymmetry of
order $5$ to be generated in the early universe, and suggests a new
mechanism for understanding the dark matter to baryon ratio.  The
penton mass is in the single eV range.  Its dominant couplings to
ordinary matter come through flavor violation.   It will be emitted
in flavor changing charged current decays of ordinary hadrons and
leptons.   The hadronic processes are enhanced by two powers of the
ratio of strong and weak fine structure constants, but even these
hadronic branching ratios for associated penton production are quite
small. There is additional chiral suppression for processes
involving only light quarks.

Gravitinos are extremely light in the Pentagon model and have no
dangerous cosmological consequences.   It is conceivable that double
gravitino exchange might be observed in short distance gravity
experiments, and that light gravitino signals can be seen at LHC.

Possible problems with the model arise in the Higgs and Singlet
sectors. It is not clear that the VEV of the singlet, $S$, is large
enough to play the role of the $\mu$ term of the MSSM.   It may be
inconsistent with perturbative unification to take the Yukawa
couplings $g_S$ and $g_T$ to be as large as they must be in order to
have acceptably large gaugino masses.   It is not clear whether the
nominal factor of $6$ between the value of electroweak VEVs  and the
scale $\Lambda_5 \sim 1.5 $ TeV, constitutes a fine tuning or little
hierarchy problem.   The small Higgs mass preferred by precision
electroweak data might also appear problematic.  The value of the
electroweak VEV obtained by setting $F_S = 0$ (which is not the
correct thing to do, since SUSY is violated) is of order $g_S
\Lambda_5$, so we might try to attribute the ratio $<H_u>/\Lambda_5$
to a small value of $g_S$.   However, setting $g_S = 0$ restores the
$Z_5$ R symmetry of the ISS state in the pure Pentagon gauge theory,
and gaugino masses will vanish in this limit.    It seems clear that
we must develop more reliable methods for computing properties of
the strongly coupled Pentagon theory, in order to assess the
severity of these tuning problems.

\section{Acknowledgments}

I would like to thank M.Dine for pointing out issues connected to
the discrete R symmetries of the ISS model, as well as discussions
of other issues connected with gauge mediation. N. Seiberg and K.
Intriligator helped me to understand the ISS paper on several
occasions. H.Haber helped me to understand the implications of the
Pentagon for precision electroweak measurements. S. Thomas pointed
out a number of important experimental implications of the model.
S.Echols and J.Jones are acknowledged for conversations relating to
baryogenesis in the Pentagon model.

This research was supported in part by DOE grant number
DE-FG03-92ER40689.




  %




\end{document}